%% file: emu_paper.tex
\RequirePackage{lineno}
\documentclass[aps,prl,preprintnumbers,showpacs,twocolumn,groupedaddress]{revtex4}
\usepackage{graphicx}
\usepackage{dcolumn}
\usepackage{bm}
\usepackage{amssymb}
\usepackage{multirow}
\usepackage{epsfig}
\usepackage{subfigure}
\usepackage{atlasphysics}
\usepackage{hyperref}

\newcommand{\vmet}{\ensuremath{\vec{\slash\kern-.7emE}_{T}}}
\newcommand{\snutau}{\ensuremath{\tilde{\nu}_{\tau}}}

\newcommand{\zp}{\ensuremath{Z'}}
\newcommand{\msnt}{\ensuremath{m_{\tilde{\nu}_{\tau}}}}
\newcommand{\mzp}{\ensuremath{m_{\zp}}}
\newcommand{\memu}{\ensuremath{m_{e\mu}}}

\begin{document}

\title{Search for a heavy particle decaying into an electron and a muon with the ATLAS detector
in $\sqrt{s} = 7$ TeV $pp$ collisions at the LHC}
\author{The ATLAS Collaboration}
\preprint{CERN-PH-EP-2011-031; submitted to PRL}
\date{\today}

\begin{abstract}
This Letter presents the first search for a heavy particle decaying
into an $e^\pm \mu^\mp$ final state in $\sqrt{s}=7$ TeV $pp$ collisions at the LHC.
The data were recorded by the ATLAS detector during 2010 and correspond
to a total integrated luminosity of 35 pb$^{-1}$. No excess above the Standard
Model background expectation is observed.
Exclusions at 95\% confidence level are placed on two representative models. In an $R$-parity
violating supersymmetric model, tau sneutrinos with a mass below 0.75 TeV are excluded,
assuming single coupling dominance and the couplings
$\lambda'_{311}=0.11$, $\lambda_{312}=0.07$.
In a lepton flavor violating model, a $\zp$-like vector boson with masses of 0.70 to 1.00 TeV and corresponding cross sections times branching ratios of 0.175 to 0.183 pb is excluded.
These results extend to higher mass RPV sneutrinos and LFV \zp s than previous constraints from the Tevatron.
\end{abstract}

\maketitle

Events with $e^\pm \mu^\mp$ ($e\mu$) in the final state, which played an important role in the discoveries
of the tau lepton and the top quark, have a clean experimental signature and low background. 
Many new physics models allow an $e\mu$ signature. 
For example, in $R$-parity violating (RPV) Supersymmetric (SUSY) models~\cite{rpv_sneutrino}
a sneutrino can decay to $e\mu$. 
Models with additional gauge
symmetry can accommodate an $e\mu$ signature through lepton flavor violating (LFV) decays of an extra gauge boson $\zp$~\cite{zprime_pap}. 
Standard Model (SM) processes that can produce an $e\mu$ signature
typically have small cross sections and the $e\mu$ invariant mass ($\memu$) lies
below the range favored for new physics signals. 
This Letter reports a search for a heavy particle decaying into the $e\mu$ final state using
data taken with the \mbox{ATLAS} detector.
The results are interpreted in terms of the production and decay of a tau sneutrino \snutau~and a $Z'$. 
Both the CDF and D0 Collaborations at the Tevatron collider have reported searches
for the RPV production and decay of the \snutau~\cite{RPVCDF1,RPVCDF2,RPVCDF3,RPVD01,RPVD02}. 
The CDF Collaboration
has also set limits on the LFV couplings as a function of the $Z'$ mass~\cite{RPVCDF2}.

The \mbox{ATLAS} detector~\cite{atlas_detector} is a multipurpose particle
physics apparatus with a forward-backward symmetric
cylindrical geometry and near $4\pi$ coverage in solid angle~\cite{coordinate}.
The inner tracking detector (ID) consists of a silicon
pixel detector, a silicon microstrip detector, and
a transition radiation tracker. The
ID is surrounded by a thin superconducting solenoid providing a
2 T magnetic field, and by a finely-segmented, hermetic calorimeter. 
The latter covers $|\eta| < 4.9$ and provides three-dimensional reconstruction 
of particle showers using lead-liquid argon sampling for 
the electromagnetic (EM) compartment followed by a hadronic compartment which
is based on iron-scintillating tiles sampling in the central region and 
on liquid argon sampling with copper or tungsten absorbers for $|\eta| > 1.7$. 
The muon spectrometer (MS) surrounds
the calorimeters and consists of three large superconducting toroids,
a system of precision tracking chambers, and detectors for triggering.

\indent The data used in this analysis were recorded in 2010 
at a center-of-mass energy $\sqrt{s} = $ 7 TeV. Application of
data-quality requirements results in a total integrated luminosity of
35 pb$^{-1}$ with an estimated uncertainty of 11\%~\cite{lumi}. 
Events are required to satisfy one of the single lepton ($e$ or $\mu$) triggers, 
which have nominal transverse momentum, $\pt$, thresholds up to 15 GeV for $e$ and 13~GeV for $\mu$. 
The trigger efficiency is measured to be 100\%, with a precision of 1\%, 
for $e\mu$ candidates containing 
two leptons with transverse momentum $\pt>20$ GeV. 

\label{sec:event_selections}
To select $e\mu$ events, the electron candidate is required to have
$\pt>20$ GeV and to lie inside the pseudorapidity regions $|\eta|<1.37$ or $1.52<|\eta|<2.47$. 
The event is rejected if the candidate cluster is located in a problematic region of the EM calorimeter. 
Electron identification and isolation requirements provide rejection against hadrons. 
A set of electron identification criteria based on the calorimeter shower shape, track quality and
track matching with the calorimeter cluster, referred to as `medium' in Ref.~\cite{electronMedium}, is
applied. In addition, a calorimeter isolation criterion
$\et^{\Delta R<0.2}/\et<0.1$ is applied, where $\et^{\Delta R<0.2}$ is defined as the transverse energy
deposited in the calorimeter within a cone of radius $\Delta R=\sqrt{\Delta \eta^2 + \Delta \phi^2}<0.2$
around the electron cluster, and \et{} is the transverse energy of the electron.

The muon candidate must be reconstructed in both the ID and the MS. 
A good match of the parameters of the ID and MS tracks is required, and the \pt{} values measured by these
two systems must be compatible. 
Furthermore, the muon candidate must have 
$\pt>20$ GeV, $|\eta|<2.4$, and be isolated in the ID with $\pt^{\Delta R<0.2}/\pt<0.1$, where
$\pt^{\Delta R<0.2}$ is defined as the sum of the \pt{} of tracks with
$\pt>1$ GeV within a cone of radius $\Delta R<0.2$ around the muon track. 

Jets are reconstructed using the anti-$k_t$ jet clustering algorithm~\cite{antiKtTopo}
with a radius parameter of 0.4. 
Only jets with $\pt>20$ GeV and $|\eta|<2.5$ are considered. 
If such a jet and an electron lie within $\Delta R=0.2$ of each other, the jet is discarded. 
Leptons are only considered if they are separated from all of the remaining jets by $\Delta R>0.4$. 
Electrons and muons are also required to be separated from each other by $\Delta R>0.2$.

The $e\mu$ candidate events are required to have exactly one electron and one muon
with opposite charge satisfying the above selection criteria. 
Furthermore, events have to contain at least one primary vertex reconstructed
with more than four associated tracks with $\pt>150$ MeV.

\indent The SM processes that can produce an $e\mu$ signature are predominantly $t\bar{t}$,
$Z/\gamma^* \rightarrow \tau \tau$, $W/Z$+jets, diboson, single top and QCD multijet events. 
Among the processes listed above, $t\bar{t}$, $Z/\gamma^* \rightarrow \tau \tau$, single top,
$WW$, $WZ$ and $ZZ$ produce electrons and muons in the final state and amount to $\sim$80\% of the
expected $e\mu$ data yield. The contributions from these processes
are estimated using Monte Carlo (MC) samples generated
at $\sqrt{s}=7$ TeV and processed with the standard chain of the \mbox{ATLAS} {\sc geant4}~\cite{geant} simulation and reconstruction~\cite{atlasMC} using the \mbox{ATLAS} MC09 parameter tune~\cite{atlasMC09}. 
The event generators used are
{\sc pythia} 6.421~\cite{pythia} ($W$ and $Z/\gamma^*$), {\sc powheg} 1.0~\cite{powheg} ($t\bar{t}$),
{\sc madgraph 4}~\cite{madgraph} ($W/Z+\gamma$), {\sc mc@nlo} 3.4~\cite{mcnlo} (single top)
and {\sc herwig 6.510}~\cite{herwig} ($WW$, $WZ$ and $ZZ$). 
The MC samples are normalized to cross sections with higher order corrections applied, as follows. 
The cross section is calculated to next-to-next-to-leading-order accuracy for $W$ and $Z/\gamma^*$~\cite{FEWZ},
next-to-leading-order plus next-to-next-to-leading-log for $t\bar{t}$
~\cite{Bonciani:1998vc}, and next-to-leading-order for $WW$, $WZ$ and $ZZ$~\cite{MCFM}. 
Single top and $W/Z+\gamma$ cross sections come from {\sc mc@nlo} and {\sc madgraph}, respectively. 
Studies of $Z/\gamma^* \rightarrow \ell \ell~(\ell = e, \mu)$ events have shown that the lepton
reconstruction and identification efficiencies, energy scale and resolution need to be adjusted
in the MC to properly describe the data. The appropriate corrections are applied
to the MC in order to improve the modelling of the backgrounds.

The processes $W/Z+\gamma$, $W/Z$+jets and pure QCD jet production
give rise to background in addition to prompt leptons, which come from $W$ and $Z$ decay. 
Jets misidentified as leptons, electrons from photon conversions, and leptons from hadron
decays (including $b$- and $c$-hadron decays) are classified as instrumental background. 
The instrumental background accounts for $\sim$20\% of the expected $e\mu$ data yield. 
The dominant component of the instrumental background comes from events with one prompt lepton 
and one jet identified as a lepton, with an additional contribution from events with two misidentified jets. 
These sources are referred to as jet instrumental background, and are estimated using data. The
background component initiated by prompt photons, referred to as the photon instrumental background, is estimated from MC.

The jet instrumental background is estimated using data in the following way. 
Two selections are defined, both requiring at least one lepton satisfying all lepton
criteria. The selections are defined by the second lepton candidate: the ``tight"
selection requires the second candidate to pass all lepton criteria, while the ``loose"
selection does not enforce the lepton isolation requirement. 
The probability for a lepton to satisfy the lepton isolation requirement is estimated by applying the
loose and tight selections on $Z/\gamma^* \rightarrow \ell\ell$ events. The probability for a jet
to satisfy the lepton isolation requirement is estimated by applying the loose and tight selections on
a sample of QCD dijet events to which
a cut on the missing transverse energy $\met<15$ GeV is applied to remove $W+$jets events. 
These two probabilities, along with numbers of events selected in
the loose and tight samples, are used to estimate the background in the final selected sample. 
The background is estimated to be $12^{+10}_{-5}$ events for candidates with one electron and
one jet satisfying the muon selection criteria, and $19^{+28}_{-7}$ events for candidates
with one muon and one jet satisfying the electron selection criteria. 
The dominant uncertainty on the jet instrumental background estimation comes from the 
\pt{} dependence of the
probability for a jet to pass the lepton isolation criterion. 
The background due to two jets satisfying the lepton requirements is estimated to be $1.3^{+5}_{-1.3}$
events from same-sign dilepton events in data with the subtraction of contributions estimated from MC for all physics processes except the QCD multijet. 
The overall jet instrumental background is found to be
$29^{+30}_{-10}$ events. 
This background level has been checked in simulated samples, which agree with the data driven estimates. 

The dominant photon instrumental background comes from the $W(\rightarrow \mu\nu)\gamma$ and
$Z(\rightarrow \mu\mu)\gamma$ processes where the photon is reconstructed
as an electron. A photon can be reconstructed as an electron if it
lies close to a charged particle track or the photon converts to $e^+e^-$ after interacting
with the material in front of the calorimeter. 
The photon instrumental background is found to be
$4.0 \pm 0.7$ events.

\indent Table~\ref{tab:data_mc_cutflow} shows the number of events selected
in data and the estimated background contributions with their uncertainties. 
A total of 160 $e\mu$ candidates are observed in data while
the expectation from SM processes is $163^{+34}_{-18}$ events. 
The dominant sources of systematic uncertainty for the SM prediction arise
from the uncertainty on the probability for a jet to satisfy the lepton isolation requirement (70\% for electron, 30\% for muon),
theoretical cross sections on
the physics background processes (5\% - 15\%),
and the integrated luminosity (11\%). 
Other systematic uncertainties from the lepton trigger, reconstruction and identification
efficiencies,
energy/momentum scale and resolution have been included and are small.

Since no excess is observed in data, limits are set for the production of 
\snutau~in RPV SUSY models and an LFV $\zp$-like vector boson. 

\newcommand{\pho}{\phantom{0}}
\begin{table}
\begin{center}
\caption{\label{tab:data_mc_cutflow} Estimated backgrounds in the selected sample, together with the observed event yield.}
\begin{tabular}{cc}
\hline
\hline
  Process & Number of events \\ \hline
 $Z/\gamma^* \rightarrow \tau \tau$  & 54$\pm$7 \\
 $t\bar{t}$  & 57$\pm$9 \\
 $WW$          & 13.4$\pm$1.7 \\
 Single top  &\pho 4.6$\pm$0.9 \\
 $WZ$          &\pho 0.79$\pm$0.11 \\
  Instrumental background &\pho 33$^{+30}_{-10}$ \\
  Total background  & 163$^{+34}_{-18}$ \\ \hline
  Data   & 160 \\
 \hline
 \hline
\end{tabular}
\end{center}
\end{table}

The RPV production of $\tilde{\nu}_{\tau}$ by
$d\bar{d}$ annihilation decaying into $e\mu$ is considered.
With the assumption of single coupling dominance~\cite{dominance},
fixing all RPV couplings but $\lambda^{'}_{311}$ ($\tilde{\nu}_{\tau}$ to $d\bar{d}$)
and $\lambda_{312}$ ($\tilde{\nu}_{\tau}$ to $e\mu$) to zero,
and that $\tilde{\nu}_{\tau}$ is the lightest supersymmetric particle,
the contributions to the $e\mu$ final state originate from the $\tilde{\nu}_{\tau}$ only.
The signal cross section
depends on the \snutau~mass ($\msnt$), $\lambda^{'}_{311}$
and $\lambda_{312}$. 
The third-generation \snutau~is considered since stringent limits exist on the electron sneutrino
and muon sneutrino~\cite{rpv_sneutrino}. 
The couplings $\lambda_{311}^{'} = 0.11$ and $\lambda_{312} = 0.07$,
compatible with the current indirect limits~\cite{rpv_sneutrino}, are chosen as a benchmark point. 

An $e\mu$ resonance can be generated in models containing a heavy neutral gauge boson with non-diagonal 
lepton flavour couplings, $Z'$~\cite{Langacker2}. Very stringent limits on the combination of the mass and 
the coupling to $ee$ and $e\mu$ of such models have been inferred from searches for rare muon 
decay~\cite{zprime_pap}. 
Using the data presented in this Letter, a limit on the production cross section times branching ratio 
to $e\mu$ can be placed on a $Z$-like vector boson. To calculate the acceptance and efficiency, the $Z'$ is assumed 
to have the same quark and lepton couplings as the SM $Z$.

MC events with \snutau~or $\zp$ decaying into $e\mu$ are generated with {\sc herwig}~\cite{herwig,herwigSig} or {\sc pythia}, respectively. 
Samples are produced with sneutrino masses ranging from 0.1 to 1 TeV, and $Z'$ masses from 0.7 to 1 TeV.

\begin{figure}[!ht]
\centering
\includegraphics[width=0.45\textwidth]{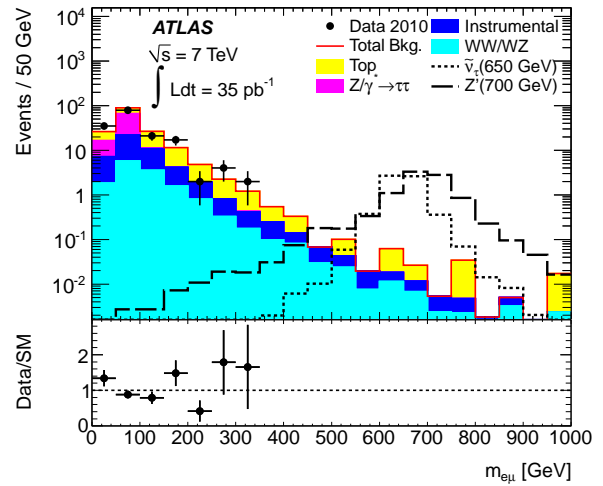}
\caption{Observed and predicted $e\mu$ invariant mass distributions.
Signal simulations are shown
for $\msnt=650$ GeV ($\lambda'_{311}=0.11$, $\lambda_{312}=0.07$) and $\mzp=700$ GeV. 
The ratio plot at the bottom includes only statistical uncertainties.}
\label{fig:data_MC_Mass}
\end{figure}

The $e\mu$ invariant mass distribution is presented in Fig.~\ref{fig:data_MC_Mass} for data,
background contributions and two possible new physics signals: a \snutau~with
$\msnt=650$ GeV and a $\zp$ with $\mzp=700$ GeV. 
The cross section is 0.31 pb for
$\msnt=650$ GeV~\cite{sneutrino_nlo} and 0.61 pb for $\mzp=700$ GeV~\cite{zp_nnlo}. 
The corresponding overall acceptance times efficiency is
55\% for \snutau~and 50\% for $\zp$.

\begin{figure}[htbp]
\centering
\includegraphics[width=0.45\textwidth]{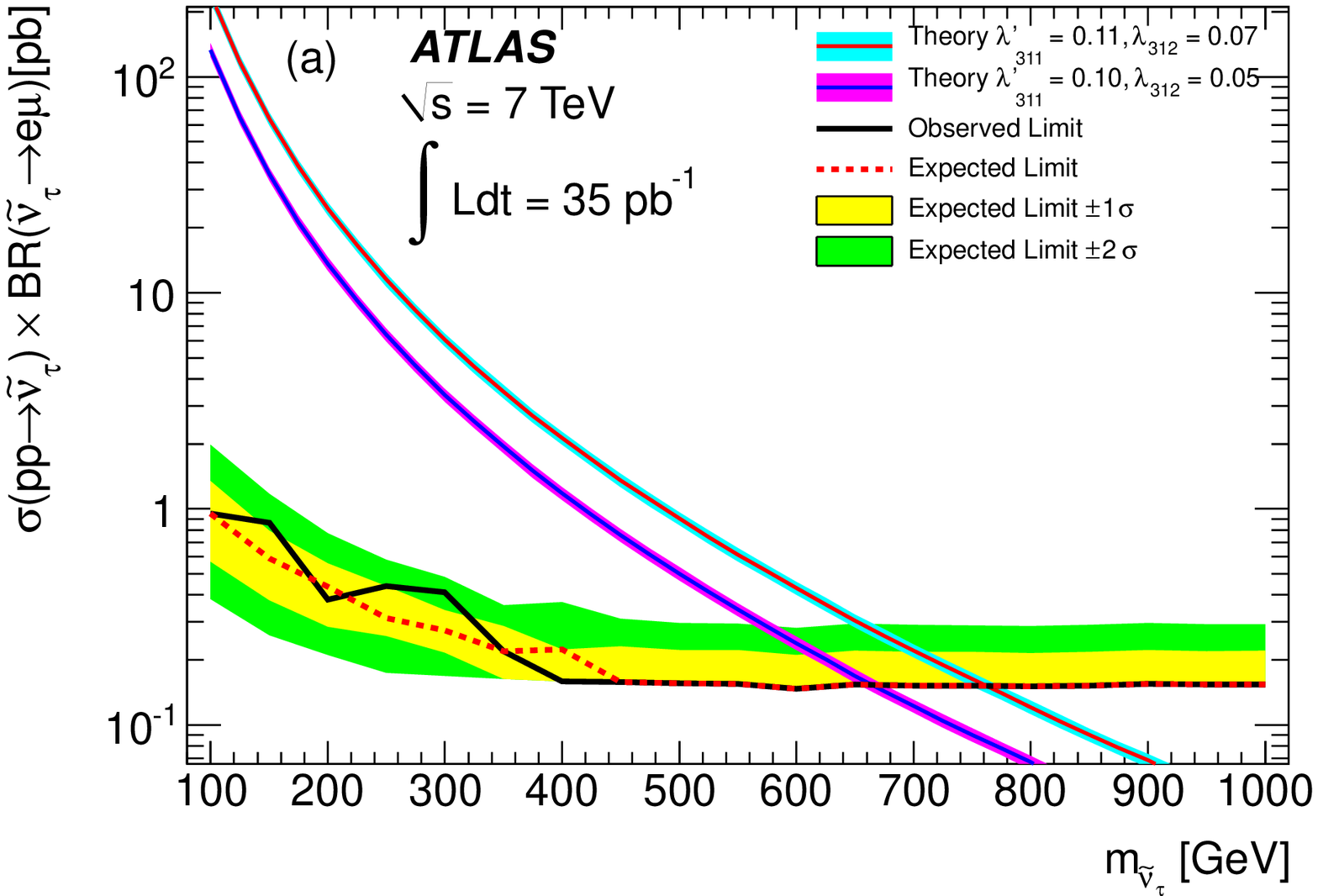}
\includegraphics[width=0.45\textwidth]{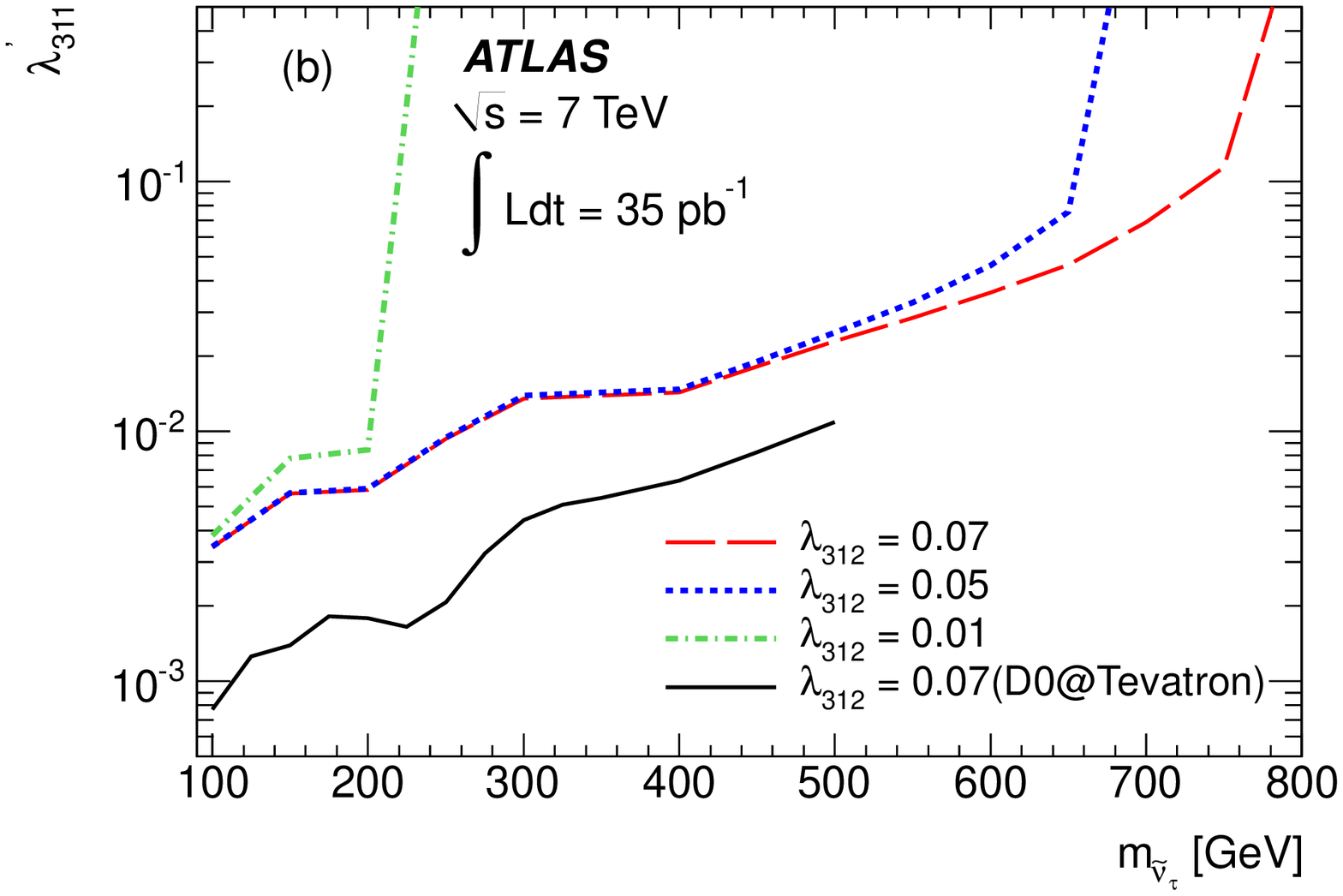}
\caption{ (a) The observed 95\% C.L. upper limits on
$\sigma(pp \rightarrow \tilde{\nu}_{\tau}) \times \text{BR}(\tilde{\nu}_{\tau} \rightarrow e\mu)$ as a function
of $\msnt$. The expected limits are also shown together with the $\pm 1$ and $\pm 2$ standard deviation
uncertainty bands. The theoretical cross sections for $\lambda'_{311}=0.11$, $\lambda_{312}=0.07$ and $\lambda'_{311}=0.10$,
$\lambda_{312}=0.05$ are also shown. 
(b) The 95\% C.L. upper limits on the $\lambda'_{311}$ coupling as a function of $\msnt$ for three values of $\lambda_{312}$. 
Regions above the three curves represent ranges of $\lambda'_{311}$ values that are excluded. 
These results are compared to the exclusion region obtained from the D0 experiment.}
\label{fig:results_parameter_limits}
\end{figure}

The $\memu$ spectrum is examined for the presence of a new heavy particle. 
For $\msnt < 500$ GeV, the search region for specific $\msnt$ is defined to be ($\msnt-3\sigma$, $\msnt+3\sigma$), where $\sigma$ is the expected $\memu$ resolution (e.g., $\sigma\simeq15$~GeV for $\msnt=$ 400~GeV). 
For higher $\msnt$, the region $\memu > 400$ GeV is used. 
The expected and observed 95\% C.L. upper limits on $\sigma(pp \rightarrow \tilde{\nu}_{\tau}) \times \text{BR}(\tilde{\nu}_{\tau} \rightarrow e\mu)$
are calculated using a Bayesian method~\cite{Bayesian} with flat prior for signal cross section as a function of
$\msnt$. 
Fig.~\ref{fig:results_parameter_limits}a shows the expected and observed limits, as a function of $\msnt$, together with the $\pm 1$ and $\pm 2$ standard deviation uncertainty bands. 
The expected exclusion limits are determined using
simulated pseudo-experiments containing only SM processes by
evaluating the 95\% C.L. upper limits for each pseudo-experiment at each 
value of $\msnt$. The median of the distribution of limits is shown as the 
expected limit. The ensemble of limits is also used to find the $1\sigma$ and $2\sigma$ envelope
of the expected limits as a function of $\msnt$. 
For a sneutrino with mass of 100 GeV (1 TeV), the limit on the cross section times branching ratio is 0.951~(0.154)~pb. 
The theoretical cross sections for $\lambda'_{311}=0.11$, $\lambda_{312}=0.07$ and $\lambda'_{311}=0.10$,
$\lambda_{312}=0.05$ are also shown. Sneutrinos with masses below 0.75~(0.65)~TeV are excluded using
$\lambda^{'}_{311}=0.11$ and $\lambda_{312}=0.07$ ($\lambda^{'}_{311}=0.10$ and $\lambda_{312}=0.05$). The results
improve on the previous CDF 95\% C.L. limit of 0.56 TeV assuming $\lambda^{'}_{311}=0.10$ and $\lambda_{312}=0.05$. 
The 95\% C.L. observed upper limits on $\lambda_{311}^{'}$ as a function of $\msnt$ are
shown in Fig.~\ref{fig:results_parameter_limits}b for three values of $\lambda_{312}$, together with the exclusion region obtained from the D0 experiment~\cite{RPVD02}. The limits derived here extend to higher mass regions.

\begin{figure}[htbp]
\centering
\includegraphics[width=0.45\textwidth]{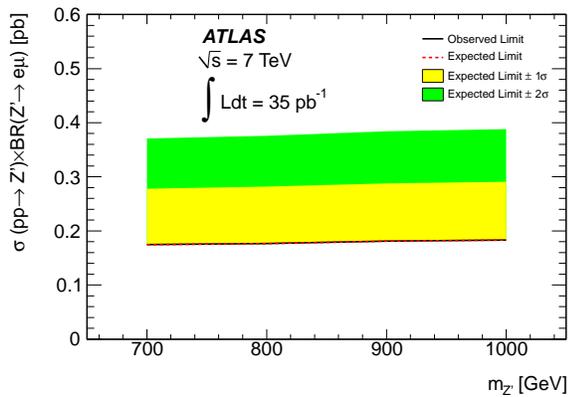}
\caption{\small The observed 95\% C.L. upper limits on $\sigma(pp \rightarrow \zp) \times \text{BR}(\zp \rightarrow e\mu)$.
The expected limits are also shown together with the $\pm 1$ and $\pm 2$ standard deviation
uncertainty bands.} 
\label{fig:results_Zprime_sigmaBR}
\end{figure}

A similar method is used to set limits on the LFV $\zp$-like vector boson, using only events with $\memu > 400$ GeV. Finding no
events in the data, the 95\% C.L. upper limits on $\sigma(pp \rightarrow \zp) \times \text{BR}(\zp \rightarrow e\mu)$ are
set, as shown in Fig.~\ref{fig:results_Zprime_sigmaBR}. 
The expected limit is the same as the observed limit because the median background event count
expectation is also zero. 
For a $\zp$ with mass of 700~GeV (1~TeV), the limit on the cross section times branching ratio is 0.175~(0.183)~pb. This result improves upon previous CDF limits by probing a higher mass range of $\zp$-like vector particles.

\indent In conclusion, a search has been performed for a heavy particle decaying into the $e^\pm\mu^\mp$ final state using
$pp$ collision data at $\sqrt{s}=7$ TeV recorded by the \mbox{ATLAS} detector.
The data are found to be consistent with the SM prediction. 
Exclusions are placed on two representative models at 95\% C.L. In an RPV SUSY
model, tau sneutrinos with a mass below 0.75 TeV are excluded,
assuming single coupling dominance and coupling values
$\lambda'_{311}=0.11$, $\lambda_{312}=0.07$. 
Higher values of the RPV coupling are also excluded as a function of $\msnt$.
In an LFV model, extra $\zp$-like gauge bosons are excluded with a cross section times branching ratio above 0.183 pb, assuming $\mzp=1$ TeV. 
These results extend to higher mass RPV sneutrinos and LFV \zp s than previous constraints from the Tevatron.

We thank CERN for the very successful operation of the LHC, as well as the
support staff from our institutions without whom ATLAS could not be
operated efficiently.

We acknowledge the support of ANPCyT, Argentina; YerPhI, Armenia; ARC,
Australia; BMWF, Austria; ANAS, Azerbaijan; SSTC, Belarus; CNPq and FAPESP,
Brazil; NSERC, NRC and CFI, Canada; CERN; CONICYT, Chile; CAS, MOST and
NSFC, China; COLCIENCIAS, Colombia; MSMT CR, MPO CR and VSC CR, Czech
Republic; DNRF, DNSRC and Lundbeck Foundation, Denmark; ARTEMIS, European
Union; IN2P3-CNRS, CEA-DSM/IRFU, France; GNAS, Georgia; BMBF, DFG, HGF, MPG
and AvH Foundation, Germany; GSRT, Greece; ISF, MINERVA, GIF, DIP and
Benoziyo Center, Israel; INFN, Italy; MEXT and JSPS, Japan; CNRST, Morocco;
FOM and NWO, Netherlands; RCN, Norway; MNiSW, Poland; GRICES and FCT,
Portugal; MERYS (MECTS), Romania; MES of Russia and ROSATOM, Russian
Federation; JINR; MSTD, Serbia; MSSR, Slovakia; ARRS and MVZT, Slovenia;
DST/NRF, South Africa; MICINN, Spain; SRC and Wallenberg Foundation,
Sweden; SER, SNSF and Cantons of Bern and Geneva, Switzerland; NSC, Taiwan;
TAEK, Turkey; STFC, the Royal Society and Leverhulme Trust, United Kingdom;
DOE and NSF, United States of America.

The crucial computing support from all WLCG partners is acknowledged
gratefully, in particular from CERN and the ATLAS Tier-1 facilities at
TRIUMF (Canada), NDGF (Denmark, Norway, Sweden), CC-IN2P3 (France),
KIT/GridKA (Germany), INFN-CNAF (Italy), NL-T1 (Netherlands), PIC (Spain),
ASGC (Taiwan), RAL (UK) and BNL (USA) and in the Tier-2 facilities
worldwide.

\onecolumngrid
\vfill
\input{atlas_authlist}
\end{document}

%% file: atlas_authlist.tex
\begin{flushleft}
{\Large The ATLAS Collaboration}

\bigskip

G.~Aad$^{\rm 48}$,
B.~Abbott$^{\rm 111}$,
J.~Abdallah$^{\rm 11}$,
A.A.~Abdelalim$^{\rm 49}$,
A.~Abdesselam$^{\rm 118}$,
O.~Abdinov$^{\rm 10}$,
B.~Abi$^{\rm 112}$,
M.~Abolins$^{\rm 88}$,
H.~Abramowicz$^{\rm 153}$,
H.~Abreu$^{\rm 115}$
% ...
\bigskip

% ...
$^{10}$ Institute of Physics, Azerbaijan Academy of Sciences, Baku, Azerbaijan\\
$^{11}$ Institut de F\'isica d'Altes Energies and Universitat Aut\`onoma  de Barcelona and ICREA, Barcelona, Spain\\
% ...
$^{48}$ Fakult\"{a}t f\"{u}r Mathematik und Physik, Albert-Ludwigs-Universit\"{a}t, Freiburg i.Br., Germany\\
$^{49}$ Section de Physique, Universit\'e de Gen\`eve, Geneva, Switzerland\\
% ...
$^{88}$ Department of Physics and Astronomy, Michigan State University, East Lansing MI, United States of America\\
% ...
$^{111}$ Homer L. Dodge Department of Physics and Astronomy, University of Oklahoma, Norman OK, United States of America\\
$^{112}$ Department of Physics, Oklahoma State University, Stillwater OK, United States of America\\
% ...
$^{115}$ LAL, Univ. Paris-Sud and CNRS/IN2P3, Orsay, France\\
% ...
$^{118}$ Department of Physics, Oxford University, Oxford, United Kingdom\\
% ...
$^{153}$ Raymond and Beverly Sackler School of Physics and Astronomy, Tel Aviv University, Tel Aviv, Israel\\
% ...
\end{flushleft}


%% file: emu_paper.bbl
\begin{thebibliography}{99}
\bibitem{rpv_sneutrino} R. Barbier {\it et al.}, Phys. Rep. {\bf 420}, 1 (2005). 
\bibitem{zprime_pap} B. Murakami, Phys. Rev. D {\bf 65}, 055003 (2002).
\bibitem{RPVCDF1} CDF Collaboration, Phys. Rev. Lett. {\bf 91}, 171602 (2003).
\bibitem{RPVCDF2} CDF Collaboration, Phys. Rev. Lett. {\bf 96}, 211802 (2006).
\bibitem{RPVCDF3} CDF Collaboration, Phys. Rev. Lett. {\bf 105}, 191801 (2010).
\bibitem{RPVD01} D0 Collaboration, Phys. Rev. Lett. {\bf 100}, 241803 (2008).
\bibitem{RPVD02} D0 Collaboration, Phys. Rev. Lett. {\bf 105}, 191802 (2010).
\bibitem{atlas_detector} ATLAS Collaboration, JINST {\bf 3}, (2008) S08003.
\bibitem{coordinate} ATLAS uses a right-handed coordinate system with its origin at the nominal interaction point (IP) in the centre of the detector and the $z$-axis along the beam pipe. The $x$-axis points from the IP to the centre of the LHC ring, and the $y$ axis points upward. Cylindrical coordinates $(r,\phi)$ are used in the transverse plane, $\phi$ being the azimuthal angle around the beam pipe. The pseudorapidity is defined in terms of the polar angle $\theta$ as $\eta=-\ln\tan(\theta/2)$.
\bibitem{lumi}ATLAS Collaboration, arXiv:1101.2185 (2011), accepted by EPJC.
\bibitem{electronMedium} ATLAS Collaboration, JHEP {\bf 1012} (2010) 060.
\bibitem{antiKtTopo} M. Cacciari, G.P. Salam, and G. Soyez, JHEP {\bf 0804} (2008) 063.
\bibitem{geant} GEANT4 Collaboration, Nucl. Instrum. Meth. A {\bf 506} 250 (2003).
\bibitem{atlasMC} ATLAS Collaboration, Eur. Phys. J. C {\bf 70}, (2010), 823.
\bibitem{atlasMC09} ATLAS Collaboration, ATL-PHYS-PUB-2010-002 (2010).
\bibitem{pythia} T. Sjostrand, S. Mrenna, and P.Z. Skands, JHEP {\bf 0605} (2006) 026.
\bibitem{powheg} P. Nason, JHEP {\bf 0411} (2004) 040; S. Frixione, P. Nason and C. Oleari, JHEP {\bf 0711} (2007) 070; S. Frixione, P. Nason and G. Ridolfi, JHEP {\bf 0709} (2007) 126.
\bibitem{madgraph} J. Alwall {\it et al.}, JHEP {\bf 0709} (2007) 028.
\bibitem{mcnlo} S. Frixione and B.R. Webber, JHEP {\bf 0206} (2002) 029; S. Frixione, E. Laenen and P. Motylinski, JHEP {\bf 0603} (2006) 092; S. Frixione {\it et al.}, JHEP {\bf 0807} (2008) 029.
\bibitem{herwig} G. Marchesini {\it et al.}, Computer Phys. Commun. 67 (1992) 465; G. Corcella {\it et al.}, JHEP {\bf 0101} (2001) 010.
\bibitem{FEWZ} K. Melnikov and F. Petriello, Phys. Rev. D {\bf 74}, 114017 (2006).
\bibitem{Bonciani:1998vc} R. Bonciani {\it et al.}, Nucl. Phys. B {\bf 529}, 424 (1998).   
\bibitem{MCFM} J. Campbell, R. Ellis and D. Rainwater, Phys. Rev. D {\bf 68}, 094021 (2003).
\bibitem{dominance} M. Chetob, Prog. Part. Nucl. Phys. {\bf 54}, 71-191 (2005).
\bibitem{Langacker2} P. Langacker, Rev. Mod. Phys. {\bf 81}, 1199 (2009).
\bibitem{herwigSig} S. Moretti {\it et al.}, JHEP {\bf 0204} (2002) 028.
\bibitem{sneutrino_nlo} S.M. Wang {\it et al.}, Phys. Rev. D {\bf 74}, 057902 (2006); S.M. Wang et al., Chin. Phys. Lett. {\bf 25}, 58 (2008).
\bibitem{zp_nnlo} R. Hamberg, W.L. van Neerven, and T. Matsuura, Nucl. Phys. B {\bf 359}, 343 (1991). 
\bibitem{Bayesian} J. Heinrich and L. Lyons, Annual Review of Nuclear and Particle Science, {\bf 57} 145 (2007).
\end{thebibliography}
